\documentclass[proceedings, preprint]{rmaa}

\usepackage{paralist}

\usepackage{psfrag,color}

\SetYear{2010}
\SetConfTitle{XIII Latin American Regional IAU Meeting}

\title{Supernova Feedback and \\ the Bend of the Tully-Fisher Relation} 

\author{
  M. E. De Rossi,\altaffilmark{1,2,3} 
  P. B. Tissera,\altaffilmark{1,2}
  and S. E. Pedrosa\altaffilmark{1,2}}

\altaffiltext{1}{Instituto de Astronom\'{\i}a y F\'{\i}sica del Espacio,
  CONICET-UBA, CC 67,  Suc 28 (1428), Ciudad Aut\'onoma de Buenos Aires, Argentina 
  (derossi@iafe.uba.ar, alternative address: mariaemilia.dr@gmail.com).}

\altaffiltext{2}{Consejo Nacional de Investigaciones Cient\'{\i}ficas y T\'ecnicas,
  CONICET, Ciudad Aut\'onoma de Buenos Aires, Argentina.}

\altaffiltext{3}{Facultad de Ciencias Exactas y Naturales, Universidad de Buenos Aires, 
  Ciudad Aut\'onoma de Buenos Aires, Argentina.}

\shortauthor{De Rossi, Tissera, \& Pedrosa}
\shorttitle{The Bend of the Tully-Fisher Relation}

\listofauthors{M. E. De Rossi, P. B. Tissera, \& S. E. Pedrosa}
\indexauthor{De Rossi, M. E.}
\indexauthor{Tissera, P. B.}
\indexauthor{Pedrosa, S. E.}

\abstract{
We have studied the origin  of the Tully-Fisher
relation by analysing  hydrodynamical simulations in a $\Lambda$CDM
universe.  We found that
smaller galaxies exhibit lower stellar masses than those predicted
by the linear fit to high mass galaxies (fast rotators), consistently with
observations.  In this model,  these trends are generated by the more
efficient action of Supernova feedback in the regulation of the star 
formation in smaller galaxies.  Without introducing scale-dependent
parameters, the model predicts that the Tully-Fisher relation
bends at a characteristic velocity of $\sim 100 \ {\rm km \ s^{-1}}$, in agreement with previous observational
and theoretical findings.
}

\resumen{
Estudiamos el origen de la relaci\'on de Tully-Fisher analizando
simulaciones  hidrodin\'amicas en un universo $\Lambda$CDM.
Encontramos que las galaxias m\'as peque\~nas exhiben masas estelares menores
que aquellas predichas por el ajuste lineal de las galaxias m\'as masivas, 
en consistencia con las observaciones.
En este modelo, estas tendencias son generadas por la acci\'on m\'as eficiente
de la retroalimentaci\'on al medio por Supernovas en la regulaci\'on de la
formaci\'on estelar de las galaxias m\'as peque\~nas.
Sin introducir par\'ametros dependientes de escala, el modelo predice
un cambio de pendiente en la relaci\'on de Tully-Fisher para una velocidad
caracter\'{\i}stica de $\sim 100 \ {\rm km \ s^{-1}}$,
en acuerdo con resultados observacionales y te\'oricos previos.}

\addkeyword{cosmology: theory}
\addkeyword{galaxies: evolution}
\addkeyword{galaxies: formation}

\begin{document}
\maketitle

\section{Introduction}
\label{sec:intro}
The Tully-Fisher relation is very important for constraining 
galaxy formation  models because it
links two fundamental properties of
galaxies such as the stellar (sTFR) or baryonic (bTFR) mass
and the depth of the  potential well.

In particular, there are observational evidences that the sTFR does not follow a
linear trend for the whole range of observed rotation velocities.
According to the results of \citet{mcg2000}, the sTFR
bends at $\sim 90 \ {\rm km \ s^{-1}}$ in such a way that smaller galaxies
have lower stellar masses than those derived from the extrapolation
of the linear fit to fast rotators.
Moreover, \citet{mcg2010} have reported that there is also
a bend in the bTFR but at a lower rotation velocity ($\sim 20 \ {\rm km \ s^{-1}}$).

\section{Simulations}
In this work, we studied the role of Supernova (SN) feedback on the shape of the sTFR
and bTFR by analysing hydrodynamical simulations in a $\Lambda$CDM universe
\footnote{${\Omega}_{\rm m}=0.3$, ${\Omega}_{\Lambda}=0.7$, ${\Omega}_{\rm b}=0.04$
and ${\rm H}_{0} = 100  \, h^{-1} \, {\rm km} \, {\rm s}^{-1} \, {\rm Mpc}^{-1}$ with $h=0.7$}.
A version of the chemical code GADGET-3 including treatments for metal-dependent
radiative cooling, stochastic star formation and SN feedback
\citep{scan2006} was employed.
The simulated volume corresponds to a cubic box of a comoving
10 Mpc $h^{-1}$ side length.  The simulation has a mass resolution of
$6 \times 10^6 M_{\odot} h^{-1}$ and $9 \times 10^5 M_{\odot} h^{-1}$ for the dark and
gas phase, respectively.

Simulated disc-like galaxies were identified following the methods describe by \citet{derossi2010}.
The mean properties of galactic systems were estimated
at the baryonic radius $R_{\rm bar}$, defined as the one which encloses 83 per cent of the
baryonic mass of the systems.
We found that the tangential velocity of these systems constitutes a good representation
of their potential well so that, for the sake of simplicity, we used the circular velocity
estimated at $R_{\rm bar}$ as the kinematical indicator for our study.

\section{Results and discussion}

The massive-end of the simulated sTFR can be fitted by a power-law of the form
$\log (M_{*} / M_{\odot} h^{-1}) = (3.68 \pm 0.09) \times \log (V / \ 100 \ {\rm km} \ {\rm s}^{-1})$ $+ (9.42 \pm 0.26)$,
in general good agreement with observations.  However, at rotation velocities below
$\sim 100 \ {\rm km \ s^{-1}}$, the sTFR becomes steep and the residuals of the linear
fit depart systematically from zero, consistently with the findings of \citet{mcg2000}. 
To analyse the role of SN feedback on the origin of the bend of the sTFR, we
run a feedback-free simulation. We found that, when 
SN feedback is suppressed from the model, the sTFR describes a linear behaviour
indicating the crucial role of SNe in the modulation of this relation.

With respect to the bTFR, both the SN feedback model and the feedback-free run show
a single slope for the relation at least for the range of velocities resolved by these simulations
($40 \ {\rm km} \ {\rm s}^{-1} < V < 250 \ {\rm km} \ {\rm s}^{-1}$).  These results
are not in disagreement with the findings of \citet{mcg2010} because they reported a bend
in the bTFR at $\sim \ 20 \ {\rm km} \ {\rm s}^{-1}$.

By analysing the simulations at higher redshifts ($0 < z < 3$), we obtained similar trends.
We have also checked that our results are robust against numerical resolution.  For further details,
the reader is referred to \citet{derossi2010}.

Finally, we explore how the SN feedback model is capable of reproducing these behaviours without
introducing scale-dependent parameters.
To analyse the impact of galactic outflows in the simulated galaxies, we defined the fraction
$f^{\rm vir}_{\rm b}$ as the ratio between the baryonic mass within the virial radius
to the one inferred from the universal baryonic fraction (${\Omega}_{\rm b} / {\Omega}_{\rm m}$).
In these simulations, $f^{\rm vir}_{\rm b}$ is within the range $0.2 \-- 0.8$ showing that for the whole sample a significant
percentage of the gas in blown away as a consequence of efficient galactic winds.  The more prominent
losses are obtained for smaller galaxies.

Moreover,  by comparing the SN feedback model with the feedback-free run, we found that SN winds 
generate an important decrease in the star formation activity of galaxies with the larger
effects in slow-rotating systems, consistently with the bend of the sTFR.
Indeed, our results suggest that the role of SN feedback on the regulation of the star formation
strongly depends on the mass of the galaxies.  In particular, we distinguish two distinct regimes
for the thermodynamical transitions of the gas phase.
In the case of smaller galaxies, the virial temperatures are lower and SN heating is more efficient
at promoting gas from a cold to a hot phase (see \citealp{scan2006} for more details about the model).
However, the cooling times of these systems are shorter than the dynamical times and
the hot gas can return to the cold phase in short time-scales.  Therefore, for slow-rotators, SN feedback
leads to a self-regulated cycle of heating and cooling strongly influencing the star formation
activity of these systems.
In the case of massive galaxies, the hot phase is established at a higher temperature and SN heating
cannot generate an efficient transition of the gas from the cold to the hot phase.  Meanwhile, the cold
gas remains available for star formation.  On the other hand, the cooling times for these galaxies get longer compared
to the dynamical times and the hot gas is able to remain in the hot phase during longer time-scales.
Hence, SN feedback is not efficient at regulating the star formation in larger galaxies.
Interestingly, in this model, the transition from the efficient to the inefficient
cooling regime for the hot-gas phase occurs at the same characteristic rotation velocity where
the sTFR bends ($\sim 100 \ {\rm km \ s^{-1}}$) and is also in agreement with previous observational
\citep[e.g.][]{mcg2000}
and theoretical works \citep[]{larson1974, ds1986}.

\section{Conclusion}
We studied the Tully-Fisher relation by performing numerical simulations in
a cosmological frame-work.  We obtained a sTFR and a bTFR in general agreement 
with observations.  The SN feedback model is able to reproduce the observed
bend of the sTFR at a characteristic velocity of $\sim 100 \ {\rm km \ s^{-1}}$ without
introducing scale dependent parameters.
We found that this characteristic velocity separates two different regimes:
slow-rotators develop a self-regulated cycle of efficient SN heating and radiative cooling
while, in fast-rotators, the cold an hot gas-phases seem to be more disconnected.
The reader is referred to \citet{derossi2010} for more details about this work.

MEDR and SEP
thank the Organizing Committee for its partial financial support to attend this meeting.
We acknowledge support from the  PICT 32342 (2005) and
PICT 245-Max Planck (2006) of ANCyT (Argentina).
Simulations were run in Fenix and HOPE clusters at IAFE and Cecar cluster at  University
of Buenos Aires.

\end{document}